\documentclass[american,aps,twocolumn,showpacs]{revtex4}
\usepackage[latin9]{inputenc}
\usepackage{dcolumn}
\usepackage{color}
\usepackage{textcomp}
\usepackage{amsmath}
\usepackage{amssymb}
\usepackage{graphicx}

\makeatletter

\usepackage{dcolumn}
\usepackage{bm}
\usepackage{wrapfig}
\usepackage{subfigure}
\usepackage{ucs}
\usepackage{lineno}
\usepackage{epsfig}
\usepackage{psfrag}\usepackage{epstopdf}
\DeclareGraphicsExtensions{.pdf,.eps,.png,.jpg,.mps}

\usepackage{babel}

\makeatother

\usepackage{babel}
\begin{document}

\title{Ground state and thermodynamic properties of spin-1/2 isosceles Heisenberg
        triangles for V$_6$-like magnetic molecules}

\author{J. Torrico$^1$ and J. A. Plascak$^{1,2,3}$ }

\affiliation{$^1$  Departamento de Física, Universidade Federal de Minas
Gerais, C. P. 702, 30123-970, Belo Horizonte-MG, Brasil}
\affiliation{$^2$ Departamento de F\'{\i}sica, Universidade Federal da Para\'{\i}ba,
Caixa Postal 5008, 58051-900, Jo\~ao Pessoa-PB, Brazil}
\affiliation{$^3$University of Georgia, Department of Physics and Astronomy,
30602 Athens-GA, USA}

\begin{abstract}
The spin-1/2 Hamiltonian for two coupled isosceles Heisenberg triangles, which is well suited
for describing the V$_6$-type magnetic molecules, is studied by exact diagonalization.
The quantum phase transition diagram, at zero temperature, is obtained as a function of
the theoretical parameters. The zero temperature magnetization is also obtained as a function
of the external magnetic field. The thermodynamic behavior of the magnetization, entropy,
susceptibility, and specific heat, as a function of temperature, are also computed and the
corresponding magnetocaloric effect analyzed for various values of the Hamiltonian parameters.
\end{abstract}

\maketitle

\section{Introduction}
\label{intro}

The study of molecular magnets has attracted the attention of the scientific community
for almost four decades now. In these systems, where the physical realizations
occur either as crystals or powders, the intramolecular interaction is much stronger than
the intermolecular interaction, making them behave mainly as an ensemble of single
independent molecules \cite{cororev,rev}. In addition, their magnetic features promise a variety
of applications in physics, magneto-chemistry, biology, biomedicine and material sciences
\cite{ses,gat,coro}, as well as in quantum computing \cite{leu,wang,wang2} and
spintronics (see Ref. \cite{cororev} and references therein).

It is well known that low dimensional magnetic systems have been broadly studied in the
literature, both experimentally and theoretically, due to the interesting magnetic and
thermodynamic properties they present at zero and finite temperatures as well. Some of these
systems have also frustrated spins due to their geometric structure (see, for instance,
\cite{Tor,galis,ekiz}). This fact makes the study of molecular magnets more attractive
still, because it turns out to be a vast field of quantum phenomena in nanosystems, in
particular zero-dimensional magnetic clusters with potential applicability in high-capacity
data storage.

An interesting property that various types of these systems exhibit is the so called
magnetocaloric effect, which consists of an isothermal change in entropy or an adiabatic
change in temperature when an external magnetic field is varied. This effect has in fact
an immense potential application \cite{tishin}. There are indeed several studies in the
magnetocaloric effect on quasi-one-dimensional spin models, such as the diamond spin chain
\cite{galisova,gal,strecka,torrico,pereira}, the antiferromagnetic triangular lattice
\cite{stre} and the octahedral chain \cite{strec}, among others.

It is thus clear from the above discussion that compounds of weakly coupled magnetic
molecules allow us to study magnetic systems at the level of nanoscale. Among the family
of molecular magnets are the various polyoxovanadates, which
contain several spin-1/2 sites originating from the vanadium ions, denoted as V$_3$, V$_6$,
V$_{12}$ and V$_{15}$ clusters, the numbers designating the corresponding quantity of vanadium ions
always occurring in triangle geometries (for more details on references about
theoretical and experimental realizations
on these compounds see \cite{kowa}). In particular,
the V$_6$ magnetic molecule is one of these systems consisting
of a pair of triangles that are weakly antiferromagnetically coupled with a strong
antiferromagnetic intratriangle coupling. This type of structure presents thus frustration
in its fundamental state, being an interesting system to study its magnetic and thermodynamic
properties \cite{muller,corti}. There are in fact two species of V$_6$ magnetic
molecule based on polyoxovanadates, which are  given by the formulae
Na$_6$H$_4$V$_6$O$_8$(PO$_4$)$_4$[(OCH$_2$)$_3$CCH$_2$OH]$_2\cdot$18H$_2$O and
(CN$_3$H$_6$)$_4$Na$_2$H$_4$V$_6$O$_8$(PO$_4$)
$_4$[(OCH$_2$)$_3$CCH$_2$OH]$_2\cdot$14\linebreak H$_2$O.

Earlier experimental and theoretical studies of both V$_6$ magnetic molecules presented above have
shown that they can be described by a spin-1/2 Heisenberg model defined on two identical
uncoupled triangles (or trimers) of spins, where in each triangle the spins interact via isotropic
antiferromagnetic exchange couplings \cite{luban}.
However, two of the three V-V interactions have exchange constants that are equal and an order of
magnitude larger than the third one, making thus a kind of an isosceles triangle in the
energy configuration. In this case, for small external magnetic field the ground state of
the molecule has a total $S=1$ spin, while for higher magnetic field the molecular spin is $S=3$.

On the other hand, additional experiments on V$_6$ compounds have further suggested that the
triangles can in fact be weakly connected through an extra superexchange interaction which is
still about 20 times smaller than the smallest triangle
interaction \cite{rouso,haraldsen}, making the molecule to behave as a hexamer.
More recently, the coupled triangles (hexamer) model with the experimental data of the exchange
interactions obtained in Ref. \cite{rouso} have been studied through exact diagonalization and
it has been shown that the molecule presents a magnetocaloric behavior and an additional magnetic
phase as a function of an external magnetic field \cite{kowa}. To be more specific, besides
the $S=1$ and $S=3$ spin phases for low and high magnetic fields, respectively, the authors
detected an extra $S=2$ spin phase. However, this new phase occurs only for a quite narrow
intermediate range of the external magnetic field.

Motivated by these previous theoretical and experimental results, we have extended the
analytical and numerical analysis of the two coupled isosceles triangles Hamiltonian to the
whole range of the exchange parameters. We will see that this rather simple magnetic molecule
actually exhibits a richness in its thermodynamic behavior which could not be forecast in
the earlier studies due to fact that just weak intertriangle interactions have been considered.

By exactly diagonalizing the Hamiltonian matrix the
entire energy spectrum and the corresponding eigenstates are obtained and the quantum phase
transitions are determined. The extra spin $S=2$ phase occurs indeed only with the presence
of the antiferromagnetic intertriangle interaction and the range of this phase depends on
the strength of this interaction. It will also be seen in the next sections that a
new phase, with zero total spin $S=0$, quite similar to an antiferromagnetic phase, is always
present for low magnetic fields. In addition, due to the degeneracy of the ground states, these
magnetic molecules present residual entropy for some range of the parameters as well as an enhanced
magnetocaloric behavior.

We would like to stress that although the experimental realizations of V$_6$
magnetic molecules have specific interaction strengths, as reported in Refs. \cite{rouso,haraldsen},
which are quite small and out of the range of the values considered in this work,
the study of the proposed isosceles triangle spin Hamiltonian for general exchange
interaction parameters are in fact quite interesting. The main reason lays on the fact that
either other new related compounds could indeed
present different stronger couplings, or even because the corresponding exchange couplings
of the known polyoxovanadates can be altered, for instance, by applying an external hydrostatic
pressure on the samples \cite{koma,wan}.

The outline of this paper is as follows. In the next section, we describe the Hamiltonian model,
present the exact analytical solution for the energy spectrum and corresponding eigenstates,
followed by the ground state phase diagram. In section \ref{thermo} we investigate
the magnetic and thermodynamic properties of the system through the study of the magnetization,
susceptibility, entropy, magnetocaloric effect and specific heat. Finally, some concluding
remarks are drawn in the last section.

\section{Hamiltonian model, eigenenergies, eigenstates and ground state phase diagram}
\label{model}

\subsection{Hamiltonian model}

\begin{figure}[h]
\includegraphics[scale=0.15]{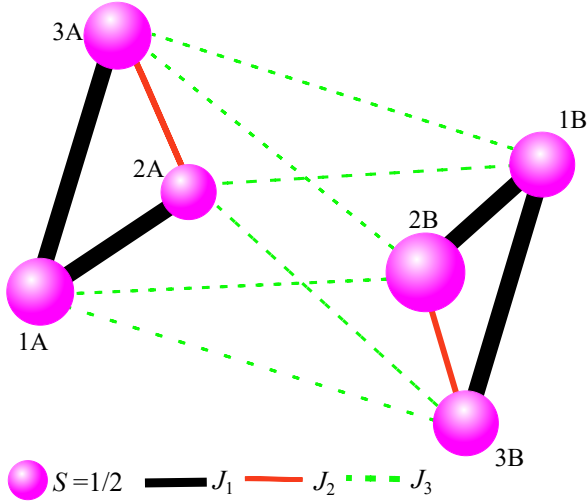}
\caption{\label{fig:1} (color online) A schematic illustration of a V$_6$ molecule described
by two triangles: one
 triangle with sites 1A, 2A, 3A, and a second triangle with sites 1B, 2B, 3B. The vanadium atoms,
 with spin-1/2, are represented by circles. The thicker lines represent the exchange interaction
 $J_1$, the thinner lines the exchange interaction $J_2$, and the dashed lines the intratriangle
 interaction $J_3$.}
\end{figure}

The V$_6$ magnetic molecule contains six vanadium ions disposed as two independent triangles,
whose sites can be denoted by 1A, 2A, 3A and 1B, 2B, 3B, respectively, as sketched in
figure \ref{fig:1}. Each vanadium has spin-1/2 and
their couplings are not symmetric but quite distorted. Spin in site 1A (1B) couples to
spins in sites 2A and 3A (2B and 3B) with the predominant exchange interaction $J_1$.
The coupling $J_2$ between sites 2A and 3A (2B and 3B) is one order of magnitude smaller
than $J_1$. This gives the molecule an isosceles triangle character under the
energy configuration point of view. Note that the superexchange couplings $J_3$ between the two
triangles are also distorted, where site 1A couples only with 2B and 3B, site 2A couples with
1B and 3B, and site 3A couples with 1B and 2B \cite{rouso,haraldsen}.
$J_3$ is still one order of magnitude smaller than $J_2$ for the vanadium V$_6$ magnetic compounds.

The corresponding Hamiltonian for these molecules can be written as
\begin{eqnarray}
  \mathcal{H}&=&J_1( \vec{S}_{1A} \cdot \vec{S}_{2A}+\vec{S}_{1A} \cdot \vec{S}_{3A} +
  \vec{S}_{1B} \cdot \vec{S}_{2B}+\vec{S}_{1B} \cdot \vec{S}_{3B}) \nonumber\\
  ~&+&J_2(\vec{S}_{2A} \cdot \vec{S}_{3A}+\vec{S}_{2B} \cdot \vec{S}_{3B})\nonumber \\
  ~&+&J_3(\vec{S}_{1A} \cdot \vec{S}_{2B}+\vec{S}_{1A} \cdot \vec{S}_{3B}
  +\vec{S}_{2A} \cdot \vec{S}_{1B}\nonumber\\
  ~&+&\vec{S}_{2A} \cdot \vec{S}_{3B}+\vec{S}_{3A} \cdot \vec{S}_{1B}+\vec{S}_{3A} \cdot \vec{S}_{2B})\nonumber\\
  ~&-&B(S_{1A}^z+S_{2A}^z+S_{3A}^z+S_{1B}^z+S_{2B}^z+S_{3B}^z),
  \label{ham}
\end{eqnarray}
where $\vec{S}_i=(S^x_i,S^y_i,S^z_i$) is the Heisenberg spin-1/2 operator with the components
$S_i^\alpha$ ($\alpha=x,y,z$) given by the Pauli spin matrices and $B$ is the external magnetic
field applied in the $z$-axis direction. All the exchange interactions $J_1,~J_2$, and $J_3$ are
positive, making each triangle a prototype example of magnetically frustrated system. In this way,
the first and second terms account for the intratriangle antiferromagnetic coupling, and the third
term accounts for the intertriangle antiferromagnetic coupling. Note that for negative value of
$J_2$ the frustation is partially removed from the molecule since each triangle would not be
frustrated.

\subsection{Eigenenergies and eigenstates}
\label{eigenen}

Taking the eigenstates of the $z$-component $S^z_i$ spin operator to spam the vectorial space of
$\cal H$, the above Hamiltonian can be represented by a $64 \times 64$ matrix that can be exactly
diagonalized by resorting to some modern technical computing system like Maplesoft\texttrademark.
The 64 eigenvalues so obtained are explicitly given in the Appendix. The corresponding eigenstates
are too lengthy to be reproduced in the Appendix and only those relevant for the quantum phase
diagram will be discussed below.

Note that the eigenenergies can be obtained for general values of the exchange interactions
$J_1,~J_2$, and $J_3$, as well as the external magnetic field $B$. It is then quite interesting
to analyze their influence on the properties of the system in a wider Hamiltonian parameter
region than that studied for only very small values of the intratriangle $J_2$ and intertriangle
interaction $J_3$ \cite{kowa}. A richer phase diagram is thus obtained where new phases, coming up
from different spin orientations, are then allowed.

After diagonalizing the Hamiltonian matrix and analyzing the corresponding eigenvalues
and eigenvectors one arrives at the following result for the most representative lower energy
spectrum of the Hamiltonian (\ref{ham}): {\it 1 - } there is
one ferromagnetic phase with total spin $S=3$; {\it 2 -} five ferrimagnetic phases with the total
spin $S=2$; {\it 3 -} four ferrimagnetic phases with the total spin $S=1$; {\it 4 -} and finally
three antiferromagnetic phases with the total spin $S=0$. Individual phases distinguish from each
other by spin arrangement as well as by the zero-temperature energies.
It worthwhile to mention that the above general trend of the lower energies has been found for
the Hamiltonian parameters considered in the following sections. For different sets of exchange
interactions the ground state picture can be slightly modified.

We present below the corresponding eigenstates of these more relevant eigenenergies grouping
them according their total spin values, since it is this quantity that gives the main magnetic
behavior of the molecule.

\subsubsection{Ferromagnetic state with $S=3$}
There is one state where all spins are aligned with the field with 
$S^z_A=S^z_B=3/2$, resulting in a molecular total spin $S=3$. We shall call this
a ferromagnetic (FM) state with energy $\mathrm{E_{FM}}$ and eigenstate $|\mathrm{FM}\rangle$
given by
\begin{align}
  \mathrm{E_{FM}}=&-3B+J_1+\dfrac{1}{2}J_2+\dfrac{3}{2}J_3,\label{EFM}\\
|\mathrm{FM}\rangle=&|+,+,+;+,+,+\rangle.\label{FM}
\end{align}
This phase is more commonly called {\it paramagnetic} phase in the literature, since all spin
components are aligned with the field.

\subsubsection{Ferrimagnetic states with $S=2$}

There are five most representative ferrimagnetic phases with total spin $S=2$, which will be
denoted by FI2$_i$ ($i=1,2,\cdots 5$). The different energies and eigenstates are given below:
\begin{align}
  \mathrm{E}_{{\mathrm{FI}2}_1}=&-2B+\dfrac{1}{2}J_1-\dfrac{1}{2}J_2,\label{EFI21}\\
  |\mathrm{FI}2_1\rangle=&\dfrac{1}{2}[|+,+,+;-,+,+ \rangle
  +|-,+,+;+,+,+\rangle\nonumber \\
  ~&-|+,+,+;+,-,+\rangle
  -|+,-,+;+,+,+\rangle],\label{FI21}
\end{align}
\begin{align}
  \mathrm{E}_{{\mathrm{FI}2}_2}=&-2B-\dfrac{1}{2}J_1+\dfrac{1}{2}J_2,\label{EFI22}\\
  |\mathrm{FI}2_2\rangle=&\dfrac{1}{2\sqrt{3}}[|+,+,+;+,-,+\rangle
  +|+,+,+;-,+,+\rangle\nonumber \\
  ~&+|+,-,+;+,+,+\rangle
  +|-,+,+;+,+,+\rangle\nonumber \\
  ~&-2|+,+,+;+,+,-\rangle
  -2|+,+,-;+,+,+\rangle],\label{FI22}
\end{align}
\begin{align}
  \mathrm{E}_{{\mathrm{FI}2}_3}=&-2B+\dfrac{1}{2}J_1-\dfrac{1}{2}J_2+J_3,\label{EFI23}\\
  |\mathrm{FI}2_3\rangle=&\dfrac{1}{2}[|+,+,+;+,-,+\rangle
  +|-,+,+;+,+,+\rangle\nonumber \\
  ~&-|+,+,+;-,+,+\rangle
  -|+,-,+;+,+,+\rangle],\label{FI23}
\end{align}
\begin{align}
  \mathrm{E}_{{\mathrm{FI}2}_4}=&-2B-\dfrac{1}{2}J_1+\dfrac{1}{2}J_2+J_3,\label{EFI24}\\
  |\mathrm{FI}2_4\rangle=&\dfrac{1}{2\sqrt{3}}[2|+,+,+;+,+,-\rangle
  +|+,-,+;+,+,+\rangle\nonumber \\
  ~&+|-,+,+;+,+,+\rangle
  -2|+,+,-;+,+,+\rangle\nonumber \\
  ~&-|+,+,+;+,-,+\rangle
  -|+,+,+;-,+,+\rangle],\label{FI24}
\end{align}
\begin{align}
  \mathrm{E}_{{\mathrm{FI}2}_5}=&-2B+J_1+\dfrac{1}{2}J_2-\dfrac{1}{2}J_3,\label{EFI25}\\
  |\mathrm{FI}2_5\rangle=&\dfrac{1}{\sqrt{6}}[|+,+,-;+,+,+\rangle
  +|+,-,+;+,+,+\rangle\nonumber \\
  ~&+|-,+,+;+,+,+\rangle
  -|+,+,+;+,+,-\rangle\nonumber \\
  ~&-|+,+,+;+,-,+\rangle
  -|+,+,+;-,+,+\rangle].\label{FI25}
\end{align}

It is interesting to notice that, in this case, both triangles are not in the eigenstates of
the squared total spin operator. In fact, the quantum state of each molecule is a mixed state
of equal statistical mixture of states with $S^z_A = 1/2$ and $S^z_B = 3/2$ or
$S^z_A = 3/2$ and $S^z_B = 1/2$, as in FI2$_1$, FI2$_3$ and FI2$_5$, and different statistical
weights as in FI2$_2$ and FI2$_4$.

\subsubsection{Ferrimagnetic states with $S=1$}

 There are four other ferrimagnetic phases with total $S=1$ spin and denoted as FI1$_i$
 ($i=1,2,\cdots 4$). The corresponding energies and eigenstates are
\begin{align}
  \mathrm{E}_{{\mathrm{FI}1}_1}=&-B-J_1-\dfrac{1}{2}J_2+\dfrac{1}{2}J_3, \label{EFI1}\\
  |\mathrm{FI}1_1\rangle=&\dfrac{1}{\sqrt{6}}[|+,+,-;+,+,- \rangle
  +|+,+,-;-,+,+\rangle\nonumber \\
  ~&+|-,+,+;+,+,-\rangle
  -|+,+,-;+,-,+\rangle\nonumber\\
  ~&-|+,-,+;+,+,-\rangle
  -|+,-,-;+,+,+\rangle],\label{FI1}
\end{align}
\begin{align}
  \mathrm{E}_{{\mathrm{FI}1}_2}=&-B-\dfrac{1}{4}J_1-\dfrac{1}{2}J_2-\dfrac{1}{4}J_3 \nonumber\\
  ~&-  \dfrac{1}{4}\sqrt{9J_1^2-10J_1J_3+17J_3^2},\label{EFI2}\\
  |\mathrm{FI}1_2\rangle=&\dfrac{1}{\sqrt{2a^2+4b^2+4}}[|+,+,+;-,+,-\rangle\nonumber \\
  ~&+a|+,+,+;-,-,+\rangle
  -b|+,+,-,+;-,+\rangle\nonumber \\
   ~&+b|+,+,-;-,+,+\rangle
  -|+,-,+;+,-,+\rangle\nonumber \\
   ~&-a|+,-,+;-,+,+\rangle
  +b|+,-,-+,+,+\rangle\nonumber \\
   ~&-b|-,+,+;+,+,+\rangle
  +|-,+,+;-,+,+\rangle\nonumber \\
   ~&-|-,-,+;+,+,+\rangle],\label{FI2}
\end{align}
where
\begin{align*}
  a=&\frac{-J_1-2J_2-J_3-\sqrt{9J_1^2-10J_1J_3+17J_3^2}}{4J_3},\\
  b=&\frac{-5J_1+2J_2+3J_3-\sqrt{9J_1^2-10J_1J_3+17J_3^2}}{4J_3},
\end{align*}
\begin{align}
  \mathrm{E}_{{\mathrm{FI}1}_3}=&-B+\dfrac{3}{4}J_1-\dfrac{3}{2}J_2-\dfrac{1}{4}J_3 \nonumber\\
  ~&-\dfrac{1}{4}\sqrt{9J_1^2-10J_1J_3+17J_3^2},\label{EFI3}\\
  |\mathrm{FI}1_3\rangle=&\dfrac{1}{\sqrt{2a^2+4b^2+4}}[|+,+,+;+,-,-\rangle\nonumber \\
  ~&+a|+,+,+;-,-,+\rangle
  -b|+,+,-;+,+,-\rangle\nonumber \\
   ~&+b|+,+,-;-,+,+\rangle
  -b|+,-,+;+,+,-\rangle\nonumber \\
   ~&-|+,-,+;+,-,+\rangle
  +b|+,-,-;+,+,+\rangle\nonumber \\
   ~&-a|-,+,+;+,-,+\rangle
  +|-,+,-;+,+,+\rangle\nonumber \\
   ~&-|-,-,+;+,+,+\rangle],\label{FI3}
\end{align}
\begin{align}
  \mathrm{E}_{{\mathrm{FI}1}_4}=&-B-\dfrac{1}{2}J_1+\dfrac{1}{2}J_2-J_3 \nonumber\\
  ~&-  \dfrac{1}{2}\sqrt{9J_1^2-10J_1J_3+5J_3^2},\label{EFI4}\\
  |\mathrm{FI}1_4\rangle=&\dfrac{1}{\sqrt{6c^2+3d^2+6}}[|+,+,+;+,-,-\rangle\nonumber \\
  ~&+|+,+,+;-,+,-\rangle
  +|+,-,+;+,-,+\rangle\nonumber \\
   ~&+|-,+,+;-,+,+\rangle
  +|-,+,-;+,+,+\rangle\nonumber \\
   ~&+|-,-,+;+,+,+\rangle
  +d|+,+,+;-,-,+\rangle\nonumber \\
   ~&+d|+,-,+;-,+,+\rangle
  +d|-,+,+;+,-,+\rangle\nonumber \\
   ~&+c|+,+,-;+,+,-\rangle
  +c|+,+,-;+,-,+\rangle\nonumber \\
   ~&+c|+,+,-;-,+,+\rangle
  +c|+,-,+;+,+,-\rangle\nonumber \\
  ~&+c|+,-,-;+,+,+\rangle
  +c|-,+,+;+,+,-\rangle],\label{FI4}
\end{align}
where
\begin{align*}
  c=&\frac{4J_1-J_2-3J_3+\sqrt{9J_1^2-10J_1J_3+5J_3^2}}{2J_3},\\
  d=&\frac{-2J_1-J_2+J_3-\sqrt{9J_1^2-10J_1J_3+5J_3^2}}{J_3}.
\end{align*}

In this ferrimagnetic phase there is statistical mixtures of states with $S_A^z=S_B^z=1/2$, with
$S_A^z=3/2$ and $S_B^z=-1/2$, and also $S_A^z=-1/2$ and $S_B^z=3/2$. The only one with equal
weights is FI1$_1$.

\subsubsection{Antiferromagnetic states with $S=0$}

 Finally, there are three antiferromagnetic phases AF$_i$ ($i=1,2,3$) with zero total spin $S=0$.
 The (field independent) energies and eigenstates are
\begin{align}
  \mathrm{E_{{AF}_1}}=&-J_1-\dfrac{1}{2}J_2-\dfrac{3}{2}J_3,\label{EAF1}\\
  |\mathrm{AF}_1\rangle=&\dfrac{1}{2\sqrt{3}}[|+,+,-;+,-,- \rangle
  +|+,-,+;+,-,-\rangle\nonumber \\
  ~&+[|+,-,-;+,+,- \rangle
  +|-,+,+;+,-,-\rangle\nonumber \\
  ~&+[|-,+,-;+,+,- \rangle
  +|-,-,+;+,+,-\rangle\nonumber \\
  ~&-[|+,+,-;-,+,- \rangle
  -|+,-,+;-,+,-\rangle\nonumber \\
  ~&-[|+,-,-;+,-,+ \rangle
  -|-,+,+;-,+,-\rangle\nonumber \\
  ~&-|-,+,-;+,-,+\rangle
  -|-,-,+;-,+,+\rangle],\label{AF1}
\end{align}
\begin{align}
  \mathrm{E_{{AF}_2}}=&-\dfrac{3}{2}J_2-\dfrac{1}{2}J_3,\label{EAF2}\\
  |\mathrm{AF}_2\rangle=&\dfrac{1}{2\sqrt{3}}[|+,+,-;+,-,-\rangle
  +|+,-,+;-,-,+\rangle\nonumber \\
  ~&+|+,-,-;+,-,+\rangle
  +|-,+,+;+,-,-\rangle\nonumber \\
  ~&+|-,+,+;-,+,-\rangle
  +|-,-,+;+,-,+\rangle\nonumber \\
  ~&-|+,+,-;-,-,+\rangle
  -|+,-,+;-,+,-\rangle\nonumber \\
  ~&+|+,-,-;+,+,-\rangle
  -|-,+,-;+,+,-\rangle\nonumber \\
  ~&-|-,+,-;-,+,+\rangle
  -|-,-,+;-,+,+\rangle],\label{AF2}
\end{align}
\begin{align}
 \mathrm{E_{{AF}_3}}=&-\dfrac{1}{2}J_1+\dfrac{1}{2}J_2-J_3\nonumber\\
  ~&-  \dfrac{1}{2}\sqrt{9J_1^2-18J_1J_3+13J_3^2},\label{EAF3}\\
  |\mathrm{AF}_3\rangle=&\dfrac{1}{\sqrt{2x^2+12y^2+6}}[x|+,+,+;-,-,-\rangle\nonumber \\
  ~&-x|-,-,-;+,+,+\rangle
  +y|+,+,-;+,-,-\rangle\nonumber \\
  ~&+y|+,+,-;-,+,-\rangle
  -y|+,+,-;+,-,-\rangle\nonumber \\
  ~&+y|-,-,+;-,+,-\rangle
  +y|-,+,+;-,+,-\rangle\nonumber \\
  ~&+y|-,+,-;+,+,-\rangle
  -y|+,-,-;+,+,-\rangle\nonumber \\
  ~&-y|+,-,-;+,-,+\rangle
  -y|-,+,+;+,-,-\rangle\nonumber \\
  ~&-y|-,+,-;+,-,+\rangle
  -y|-,-,+;+,+,-\rangle\nonumber \\
  ~&-y|-,-,+;-,+,+\rangle
  +|+,-,-;-,+,+\rangle\nonumber \\
  ~&+|-,+,-;-,+,+\rangle
  +|-,-,+;+,-,+\rangle\nonumber \\
  ~&-|+,+,-;-,-,+\rangle
  -|+,-,+;-,-,+\rangle\nonumber \\
   ~&-|-,+,+;-,-,+\rangle],\label{AF3}
\end{align}
where
\begin{align*}
  x=&\frac{2J_1+J_2+J_3-\sqrt{9J_1^2-18J_1J_3+13J_3^2}}{2(J_1-J_3)},\\
  y=&\frac{2J_1-J_2-3J_3+\sqrt{9J_1^2-18J_1J_3+13J_3^2}}{2(2J_1+J_3)}.
\end{align*}

AF$_1$ and AF$_2$ are equal statistical mixtures of states $S_A^z=1/2$ and $S_B^z=-1/2$ and the
inverse $S_A^z=-1/2$ and $S_B^z=1/2$. State AF$_3$ has different weights with the additional states
$S_A^z=3/2$ and $S_B^z=-3/2$, and $S_A^z=-3/2$ and $S_B^z=3/2$.

 \subsubsection{Brief overview of the entire energy spectrum}

\begin{figure}[h]
\includegraphics[scale=0.45]{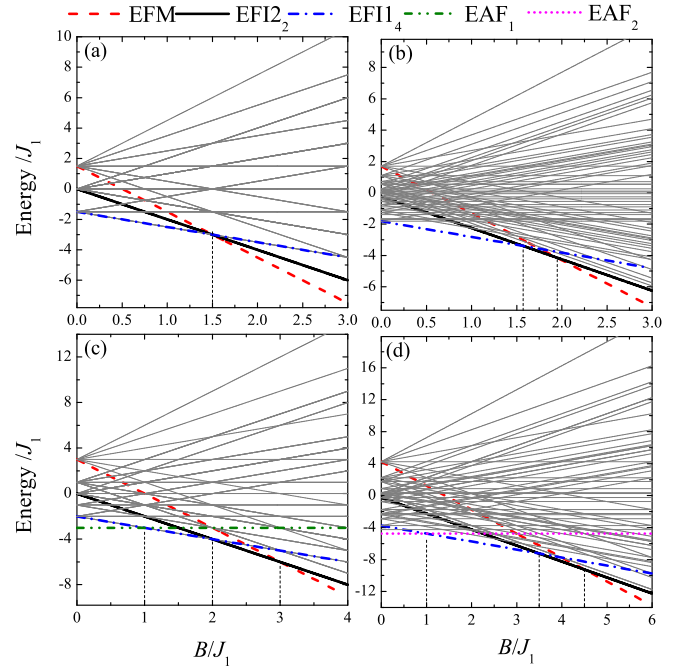}
\caption{\label{fig:2} (color online)
 Energy spectrum, parametrized by the exchange interaction $J_1$, as a
 function of the external magnetic field $B/J_1$, for some values of the exchange interactions
 $J_2$ and $J_3$. We have in (a) $J_2/J_1=1$ and $J_3/J_1=0$; in (b) $J_2/J_1=0.5$ and $J_3/J_1=0.3$;
 in (c) $J_2/J_1=1$ and $J_3/J_1=1$; and in (d) $J_2/J_1=0.5$ and $J_3/J_1=2$.
 The thicker lines gives the lowest most important energies for determining the ground state.}
\end{figure}

In Fig. \ref{fig:2} it is shown the complete spectrum of the energies as a function of the external
magnetic field for some values of the exchange interactions. All quantities are parametrized by
$J_1$. The different number of lines in each graph reflects the fact the for some particular values
of the exchange interactions the degree of the degeneracy is different. For instance, in Fig.
\ref{fig:2}(a) we have $J_3=0$, meaning that both triangles are decoupled and the spectrum is
dramatically reduced, even for zero external field. In this case, for $J_2/J_1=1$ and $B/J_1<1.5$ the stable phase is the
ferrimagnetic FI1$_4$ phase with total spin $S=1$, while for $B/J_1>1.5$ the stable phase is
the ferromagnetic FM with $S=3$. This result is completely equivalent to that obtained in Ref.
\cite{luban} when one has two independent triangles Hamiltonian for the V$_6$ magnetic molecules.

In Fig. \ref{fig:2}(b) we have $J_2/J_1=0.5$ and we consider an even weaker coupling between the
triangles $J_3/J_1=0.3$. We can see that the intertriangle interaction induces a new ferrimagnetic
phase with molecular spin $S=1$ in a small region between the external fields $1.57<B_0/J_1<1.95$, where
the ferrimagnetic state FI2$_2$ becomes the ground state with energy given by Eq. \eqref{EFI22}.
The weaker the interaction $J_3$ the smaller the field region of the stability of the FI2$_2$ phase,
which eventually disappears for $J_3=0$ [in fact, this phase becomes identical to the ones with
$S=3$ and $S=1$ at this point in a multiphase point, as can be seen in Fig. \ref{fig:2}(a)].
This is the reason in Ref. \cite{kowa} a very small range of the external field for this phase
has been obtained, since the intertriangle interaction was taken very small. However, by
increasing the intertriangle interaction, not only the range of
the FI2$_2$ phase increases, but also one of the antiferromagnetic phases (AF$_1$), with total
spin $S=0$, becomes stable for smaller external fields, as shown in Fig. \ref{fig:2}(c) for
$J_2/J_1=1$ and $J_3/J_1=1$. This zero spin phase persists by still increasing the $J_3$
interaction, but the antiferromagnetic AF$_2$ phase becomes stable instead, as depicted in
Fig. \ref{fig:2}(d) for $J_2/J_1=0.5$ and $J_3/J_1=2$.

\subsection{Ground state phase diagram }
\label{phasdia}

\begin{figure}[h]
\includegraphics[scale=0.44]{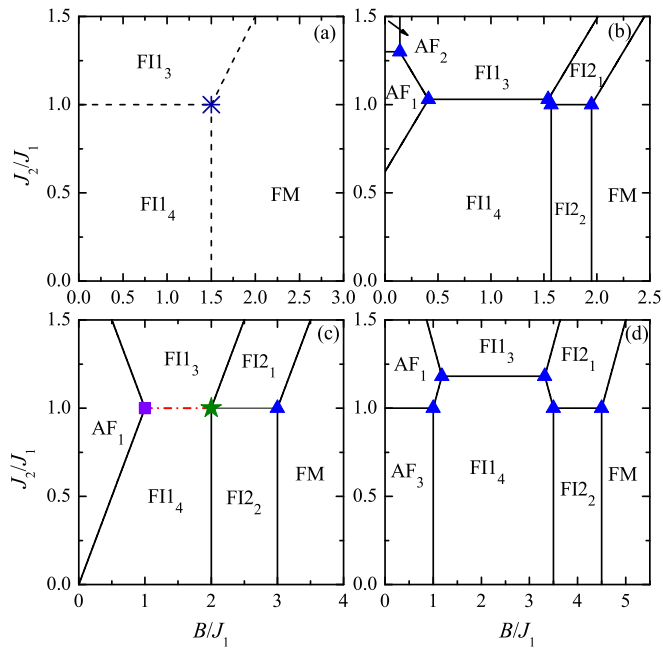}
\caption{\label{fig:3}  (color online) Ground-state phase diagram in the $B/J_1$ versus $J_2/J_1$
 plane. The full lines represent transitions where two phases coexist, the dashed-dotted line
 where three phases coexist (triple line), and along the dashed lines one has four phases
 coexisting (quadruple line).
 Accordingly, the triangles are triple points, the square is a quadruple point, the star is a
 quintuple point, and the asterisk represents a nonuple point (nine phases coexisting). The labels
 describe the corresponding ordering according to the text. We have, in increasing value of $J_3$:
 (a) $J_3/J_1=0$; (b) $J_3/J_1=0.3$; (c) $J_3/J_1=1$; and (d) $J_3/J_1=2$.
 }
\end{figure}

By expanding the analysis done above to other values of the exchange interactions,
it is possible to draw the ground state phase diagram
by seeking the lowest energy as a function of the Hamiltonian parameters.
Fig. \ref{fig:3} shows the corresponding phase diagrams in the $B/J_1$ versus $J_2/J_1$ plane
for different values of the intertriangle interaction $J_3/J_1$.
In Fig. \ref{fig:3}(a), for $J_3=0$ (decoupled triangles), we have only three different ground states,
two ferrimagnetic phases, namely FI1$_3$ and FI1$_4$, and one ferromagnetic phase FM. From the
equations of the previous subsection it is easy to see that the transition line between phases
FI1$_3$ and FI1$_4$ at $J_2=J_1$ is indeed a quadruple line, since the energy of all four
ferrimagnetic phases with total spin $S=1$ become equal in this case (the degeneracy is broken
when $J_2/J_1$ is different from 1). Similarly, quadruple lines are obtained
for $J_2/J_1<1$, with the coexisting FM, FI1$_4$, FI2$_2$, and FI2$_4$ phases, and for $J_2/J_1>1$
phases FM, FI1$_3$, FI2$_1$, and FI2$_3$ coexist. The meeting point of these three lines turns
out to be a nonuple point, where nine phases have the same energy. As both triangles of the
magnetic molecule are decoupled, a simpler phase diagram for each one of them could be obtained.
However, it is worthwhile to have this highly degenerated phase diagram in order to better
understand the limiting case of the
molecule when $J_3\rightarrow 0$. It is also interesting to see that this magnetic molecule
should present a residual entropy, given by the natural logarithm of the degeneracy of the
ground state. This feature will be more clearly seen in the next section.

The degeneracy present in the phase diagram is partly broken when the intertriangle interaction
$J_3$ is switched on. Fig. \ref{fig:3}(b) shows the corresponding phase diagram for $J_3/J_1=0.3$.
Only two-phase coexistence lines and triple points are obtained. For $J_3/J_1=1$ an additional
triple line with coexisting FI1$_2$, FI1$_3$, and FI1$_4$  phases appears, with a quadruple and
a quintuple point at its ends, as shown in Fig. \ref{fig:3}(c). For still greater values of the
intertriangle interaction, $J_3/J_1=2$ in Fig. \ref{fig:3}(d),
again only two-phase coexistence lines and triple points are obtained.

It should be noticed that the intertriangle interaction, in all diagrams depicted in Fig. \ref{fig:3},
induces an antiferromagnetic phase with zero total spin for small values of the external field.
In fact, in Fig. \ref{fig:3}(a), only for $B=0$ the antiferromagnetic phase AF$_3$ has the same energy
as phase FI1$_4$ for $J_2/J_1<1$, while for $J_2/J_1>1$ AF$_2$ and FI1$_3$ have the same energy.

The vertical lines, in the bottom region of the phase diagrams of Fig. \ref{fig:3}, extend to
smaller and eventually any negative values of $J_2$.
Note that negative values of the exchange coupling $J_2$ allow ferromagnetic
exchange interactions between sites 2 and 3 (see Fig. \ref{fig:1}). As a result, there is a
break in the frustration present in each triangle, although the frustration still
persists in the whole molecule due to the intertriangle antiferromagnetic interaction. One
should say that with negative values of $J_2$ the molecule becomes {\it less frustated}.
These vertical lines can be
located at particular values of critical external magnetic field $B_c$ (although improperly
termed critical field in earlier works of V$_6$ magnetic molecules, we will use this same
nomenclature for this field).  The corresponding values of $B_c$ are given in Table
\ref{tab1} for the different transition lines shown in Fig. \ref{fig:3}.
\begin{table}[h!]
\setlength{\arrayrulewidth}{0.4mm}
\setlength{\tabcolsep}{1pt}
\renewcommand{\arraystretch}{2.4}
\centering
\caption{Critical external magnetic field $B_c$ for the vertical transition lines in the
 bottom region of the phase diagrams shown in Fig. \ref{fig:3}.}
\label{tab1}
\begin{tabular}{|c|c|}
  \hline
  Transition  & External field  $B_c$  \\
\hline \hline
  FM$\leftrightarrow$ FI1$_4$ & $\dfrac{3}{4}J_1+\dfrac{5}{4}J_3
  +\dfrac{1}{4}\sqrt{9J_1^2-10J_1J_3+5J_3^2}$ \\ \hline

  AF$_1$$\leftrightarrow$FI1$_4$ &  $\dfrac{1}{2}J_1+J_2+\dfrac{1}{2}J_3-
  \dfrac{1}{2}\sqrt{9J_1^2-10J_1J_3+5J_3^2}$
\\ \hline

  FI2$_2$$\leftrightarrow$FI1$_4$ &
  $J_3+\dfrac{1}{2}\sqrt{9J_1^2-10J_1J_3+5J_3^2}$\\ \hline

  FI2$_2$$\leftrightarrow$FM &
  $\dfrac{3}{2}J_1+\dfrac{3}{2}J_3$\\ \hline

  AF$_3$$\leftrightarrow$FI1$_4$ &
  $\dfrac{1}{2}\sqrt{9J_1^2-18J_1J_3+13J_3^2}$\\
   & $-\dfrac{1}{2}\sqrt{9J_1^2-10J_1J_3+5J_3^2}$ \\ \hline
\end{tabular}
\end{table}

\begin{figure}[h]
\includegraphics[scale=0.45]{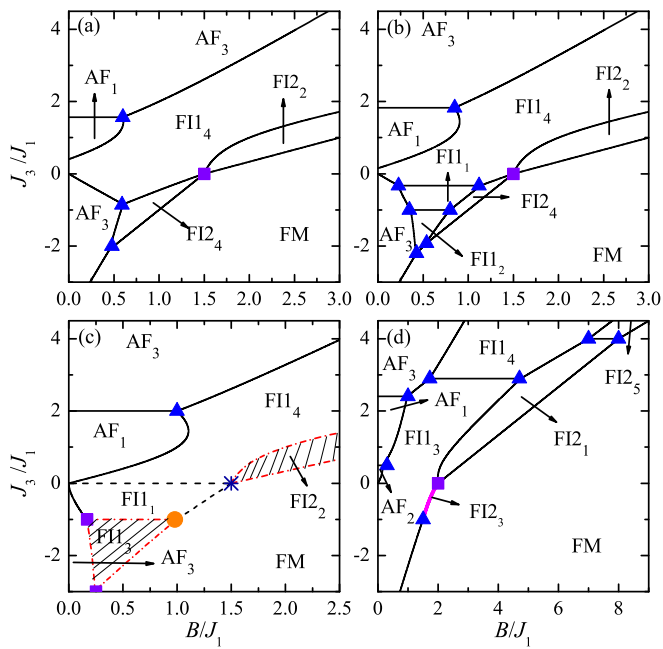}
\caption{\label{fig:4} (color online) Ground-state phase diagram in the $J_3/J_1$ versus $B/J_1$
 plane, for different values of $J_2$. The lines, symbols and geometric shapes have the same
 meaning as in Fig. \ref{fig:3}, with the additional sextuple point represented by the full circle
 and a delimited region of two phases coexistence, given by the ruled areas. In increasing values
 we have: in (a) $J_2/J_1=0.5$; (b) $J_2/J_1=0.8$; (c) $J_2/J_1=1$; and (d) $J_2/J_1=1.5$.
 In (d) one has a very narrow region, represented in this case by just a thicker line, where the phase
 FI2$_3$ is stable. }
\end{figure}

Figs. \ref{fig:4}(a)-(d) show the phase diagrams in the $B/J_1$ versus $J_3/J_1$ plane
for different values of the intratriangle interaction $J_2$. It is instructive to see
that now, instead of straight transition lines, we have some curves due to the square
root dependence of some energies on $J_3$. For fixed values of $J_3$, the square roots
appearing in all energies become constant, since they strictly depend on the ratio
$J_3/J_1$. Accordingly, only linear behavior for the transition lines are obtained in
Figs. \ref{fig:3}. There are, however, some new features in the phases topology in this
field versus intratriangle interaction that has not been seen in the phase diagrams of
Figs. \ref{fig:3}, namely an additional sextuple point, where the six phases
FM, FI1$_1$, FI1$_2$, FI1$_3$, FI2$_3$, and FI2$_4$ coexist, and an extended region of two phases
coexistence (not a line or point), as is depicted in \ref{fig:4}(c). In the ruled area
FI1$_3$, this phase coexists with FI1$_2$, whereas in the ruled area FI2$_2$, this phase
coexists with FI2$_1$ phase.

\section{Magnetic and thermodynamic properties}
\label{thermo}

Let us now turn our attention to the thermodynamic properties of these V$_6$-like molecules
in the extended region of the Hamiltonian parameters. As the intermolecular interactions
are quite weak, one can think of this system as composed of an assembly of $N$ independent
single molecules. As a result, the total Hamiltonian is given by a sum of $N$ Hamiltonians
as defined in Eq. \eqref{ham}, and the
corresponding partition function is a product of $N$ partition functions for each molecule. In
the thermodynamic limit, all the thermal properties (per molecule) of the whole system is thus
obtained from the partition function of one single molecule, which can be explicitly written as
\begin{eqnarray}
  \mathcal{Z}=\sum_{i=1}^{64}\mathrm{e}^{-\beta \varepsilon_i},
  \label{z}
\end{eqnarray}
where $\beta=1/k_BT$, $k_B$ is the Boltzmann constant, $T$ is the absolute temperature,
and $\varepsilon_i$ are all the Hamiltonian \eqref{ham} eigenvalues, which are described
in the Appendix.  The corresponding free energy per molecule is
\begin{eqnarray}
  F=-k_BT\ln(\mathcal{Z}).
\end{eqnarray}

The magnetothermodynamic quantities per molecule, such as magnetization $M$ and
susceptibility $\chi$, entropy $\mathcal{S}$ and specific heat $C$, can thus be
calculated from the well known thermodynamic relations
\begin{eqnarray}
  M&=&-\left(\dfrac{\partial F}{\partial B}\right)_T, \quad
  \chi=\left(\dfrac{\partial M}{\partial B}\right)_{T},\quad \\
  \mathcal{S}&=&-\left(\dfrac{\partial F}{\partial T}\right)_B, \quad
  C=T\left(\dfrac{\partial S}{\partial T}\right)_B.
\end{eqnarray}

Another interesting quantity that can be studied in this system is the so called magnetocaloric
effect, which is defined by the adiabatic temperature change, or the isothermal entropy change, as
the  external magnetic field is varied. This effect is can be quantified by the following relation
\begin{eqnarray}
  \left(\frac{\partial T}{\partial B}\right)_S=-\frac{(\partial S/ \partial B)_T}
  {(\partial S/ \partial T)_B}.
  \label{mceq}
\end{eqnarray}

In the next subsections a detailed study of these properties is presented for several values
of the Hamiltonian parameters. The partition function, given by Eq. \eqref{z}, is thus
numerically obtained from the corresponding energy spectrum transcribed in the Appendix.
In all figures below we have, for simplicity, taken $k_B=1$ and the
corresponding temperature has been parametrized by $T/J_1$ (reduced temperature).

\subsection{Magnetization}
\label{mag}

\begin{figure}[h]
\includegraphics[scale=0.45]{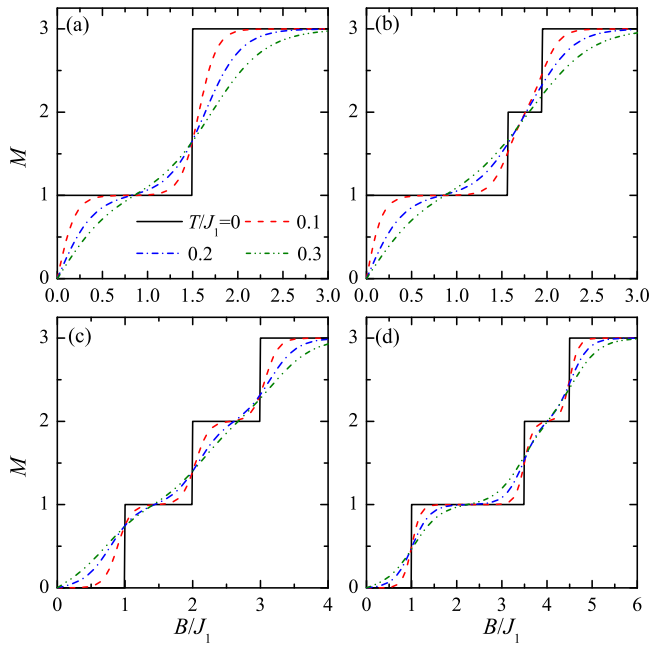} \caption{\label{fig:5} (color online)
 Total molecular magnetization $M$ as a
 function of the magnetic field $B/J_1$ and different values of temperature. In (a) $J_2/J_1=1$
 and $J_3/J_1=0$; in (b) $J_2/J_1=0.5$ and $J_3/J_1=0.3$; in (c) $J_2/J_1=1$ and $J_3/J_1=1$;
 in (d) $J_2/J_1=0.5$ and $J_3/J_1=2$.}
\end{figure}

Fig. \ref{fig:5} illustrates the total magnetization $M$, as a function of the magnetic field
$B/J_1$, for different reduced temperatures $T/J_1$ and various values of the exchange interactions.
The exchange interactions have been chosen in such a way that one has all possible transitions
from the different total molecular spin $S$, ranging from zero to three. For clarity, the label
of the phases has been omitted in Fig. \ref{fig:5}. Note that when the molecules
have a nonzero magnetization at $T=0$, as in Figs. \ref{fig:5}(a) and (b), the temperature
totally breaks
the order of the molecular spins. This reflects the fact that each molecule, being a zeroth
dimensional quantum system, is indeed equivalent to a one-dimensional classical spin model,
which in turn has no phase transition at finite temperatures.

It is clearly seen in all figures that the ground state plateaux are smoothen as soon as the
temperature is greater than zero. Consequently, the magnetization has no jumps and no transitions
take place as the external field changes. However, for low temperatures, the system keeps a kind
of {\it memory} of the behavior at zero temperature, either along the field axis or along the
magnetization axis. For instance, in Fig. \ref{fig:5}(a), for uncoupled triangles where $J_2/J_1=1$
and $J_3/J_1=0$, at $B/J_1=B_c/J_1=1.5$ and $B/J_1\sim 0.8$ the magnetization is almost independent
of the temperature, since all curves cross at almost the same value $M\sim 1.5$
and $M\sim1$, respectively. This behavior has already been experimentally observed in the two
species of polyoxovanadate-based magnetic molecules \cite{luban}. It should be stressed that this
independence is only for low temperatures, about the experimental value 20K. In Fig. \ref{fig:5}
the values of $T/J_1$ up to 0.3 is
equivalent to the experimental temperatures for the considered exchange couplings. For higher values
of $T$ the curves are much farther from the ground state plateaux. By looking at Figs.
\ref{fig:5}(b)-(d) we can notice that this same trend occurs at the molecular total spin even
for greater values of the intertriangle interaction. In these cases, there will be more values
of the magnetic field where the magnetization is independent of (low) temperature.

\begin{figure}[h]
\includegraphics[scale=0.45]{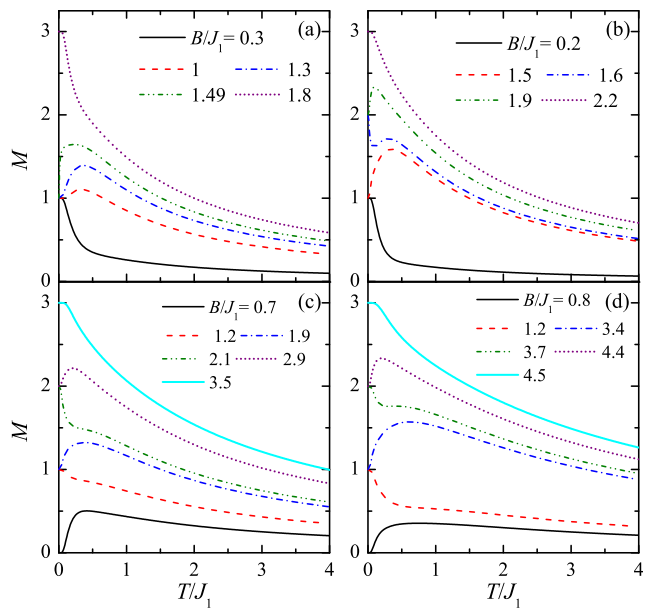} \caption{\label{mxt} (color online)
 Total molecular magnetization $M$ as a
 function of temperature $T/J_1$ for different values of the external fields. The exchange
 interactions in each graphic are the same as those in Fig. \ref{fig:5} and the fields
 are chosen so the behavior of the magnetizations are different in the low temperature region.
 }
\end{figure}

Another interesting feature that can be observed in  Fig. \ref{fig:5} is that for constant
external fields, there are regions where the magnetization of the molecule increases by
increasing the temperature, whereas in other regions the magnetization decreases with
temperature, the latter behavior being more expected from the thermodynamic point of view.
Fig. \ref{mxt} shows examples of the molecular magnetization $M$ as a function of temperature
for some external magnetic fields. The magnetic fields have been selected in different regions
of the graphics given in Fig. \ref{fig:5}. The molecular magnetization, in all cases,
eventually goes to zero at high temperatures, as expected.

The behavior of the molecular magnetization at low temperatures can be understood by looking
at the spectrum of the molecule as shown in Fig. \ref{fig:2}. For some given values of the
external field, the excited states are those having higher values of the total molecular
spin. Thus, increasing the temperature from zero those excited states are populated, which in
turn increases the magnetization. For other values of the external field the excited states
are formed by those having smaller values of the total molecular spin so the increase in
temperature actually decreases the magnetization. Eventually, in either case, the
magnetization goes to zero at high temperatures. It should be stressed that this behavior
has also been experimentally observed in the polyoxovanadate-based magnetic molecules
\cite{luban}.

\subsection{Susceptibility}
\label{sus}

\begin{figure}[h]
\includegraphics[scale=0.44]{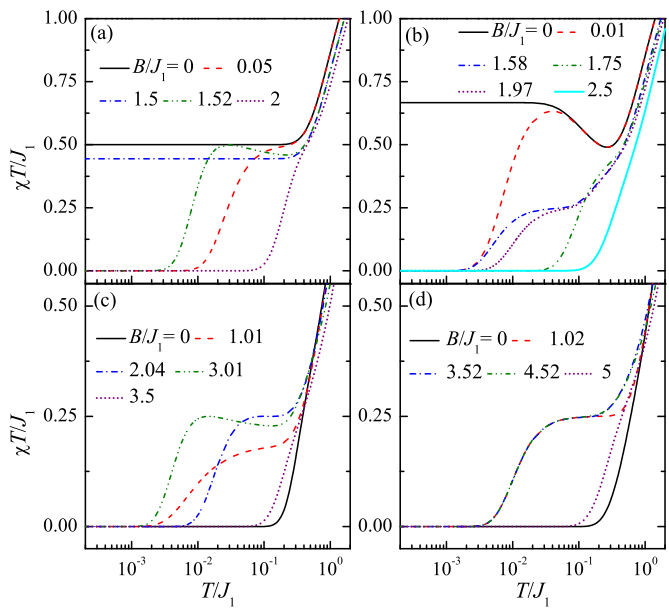} \caption{\label{fig:6} (color online) Susceptibility times
 temperature, in units of $J_1$, $\chi T/J_1$, as a function of temperature $T/J_1$ (logarithmic scale), for
 different values of the magnetic field $B/J_1$. The exchange interactions in (a)-(b) are the same as
 in Fig. \ref{fig:5}.}
\end{figure}

The susceptibility times temperature $\chi T/J_1$, as a function of temperature $T/J_1$
(in logarithmic scale to a better view of the whole range of $T$), for different magnetic fields
$B/J_1$ and exchange interactions $J_2/J_1$ and $J_3/J_1$, is illustrated in Fig. \ref{fig:6}.
At high temperatures, the susceptibility turns out to be the same, independent of the
external magnetic field, as expected. Also expected is the behavior at zero temperature,
where the plateaux in Fig. \ref{fig:5} for the total magnetic spin results in a zero
magnetic susceptibility. On the
other hand, at the critical fields, $\chi$ goes to infinity as the inverse of temperature.
However, this is not a kind of quantum phase transition, rather a paramagnetic behavior
following the Curie law. As a matter of comparison, in the one-dimensional Ising model
$\chi T\rightarrow\infty$, meaning that at zero temperature the system is indeed critical
\cite{baxter}.

We can see in Figs. \ref{fig:6} that for magnetic fields just above the smaller critical ones,
$\chi T$ increases from zero, reaches a maximum and then a round minimum is formed before
saturating to its high temperature behavior. However, for higher values of the critical field,
just a shoulder is present. The round minimum is also kind of suppressed by increasing the
$J_3$ interaction even for smaller values of critical fields. It is interesting to stress
that the qualitative behavior of the minimum shown in Fig.\ref{fig:6}(a) has indeed been
experimentally observed for the V$_6$ compounds in ref. \cite{haraldsen}. The appearance of such
round minimum in the temperature dependence of $\chi T$ is a typical feature of quantum
ferrimagnets, as has been discussed in Refs. \cite{strecka1,strecka2}. In the present case
it is enhanced close to the critical fields occurring at zero temperature.

\subsection{Entropy}
\label{entro}

\begin{figure}[h]
\includegraphics[scale=0.45]{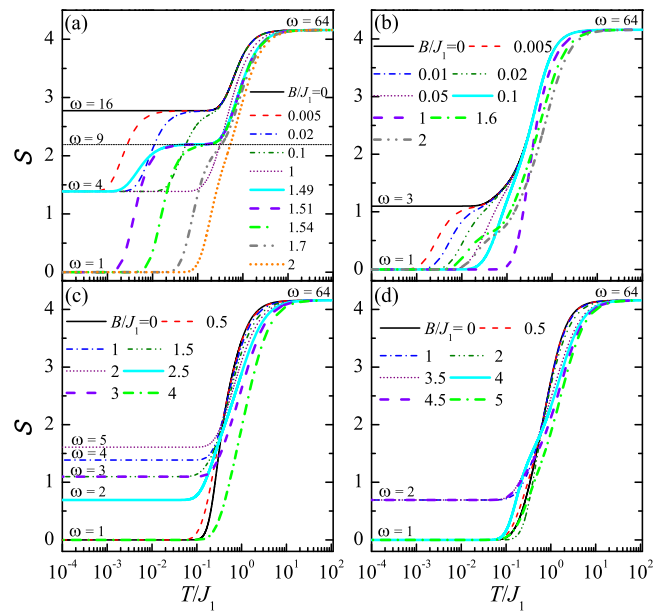}
\caption{\label{fig:7} (color online) Entropy $\mathcal{S}$ as a function of
 temperature $T/J_1$ (in logarithmic scale) and different values of the magnetic field $B/J_1$. At low
 temperatures, $\omega$ gives the degeneracy of the ground state.
 The exchange interactions in (a)-(b) are the same as those in Fig. \ref{fig:5}.}
\end{figure}

First, let us analyze the behavior of the entropy $\mathcal{S}$, as a function of temperature
$T/J_1$, as shown in Fig. \ref{fig:7} for several values of the external magnetic field and
exchange interactions. As $k_B$ has been set to unity, in some special limits the entropy
will be just given by $\mathcal{S}=\ln(\omega)$, with $\omega$ being the number of states at
that limit. For instance, we can easily see that at high temperatures, where all states are
equally probable to occur, all curves go to the limit $\mathcal{S}=4.1589$, where $\omega=64$.
On the other hand, at low temperatures, depending on the value of the Hamiltonian
parameters, different residual entropies are present in the molecule. The degree of
degeneracy, here given by $\omega$, is specified by the number of coexisting phases present at
$T=0$, and are explicitly written in Fig. \ref{fig:7}. Several examples, for some particular
values of the field and exchange interactions, are depicted in Figs. \ref{fig:3} and
\ref{fig:4}. Apart from infinite temperature, the highest degenerated state occurs at zero
field for non-interacting isotropic triangles ($J_3=0$ and $J_2=J_1$) with $\omega=16$
(as this happens along the $J_2/J_1$ axis, it has not been shown in Fig. \ref{fig:3}(a)
for questions of clarity).

It is interesting to note in Fig. \ref{fig:7}(a) that for $B/J_1=0.005$ and 0.02, before
reaching the high temperature behavior, the entropy
starts at $\mathcal{S}=2\ln(2)$ and quickly jumps to $\mathcal{S}=4\ln(2)$ at low temperature.
Conversely, for $B/J_1=1.51$ and 1.54, the system has zero entropy at its ground state but
at low enough temperatures it quickly jumps to the entropy value of $\mathcal{S}=2\ln(3)$
before reaching the high temperature behavior. The
increase of the intertriangle interactions suppresses this intermediate jump behavior of the
entropy, where only a small shoulder is still present in Fig. \ref{fig:7}(b).
However, as we will see right below, these entropy plateaux are indeed an important ingredient
for the presence of magnetocaloric effects in these magnetic molecules.

\begin{figure}[h]
\includegraphics[scale=0.4]{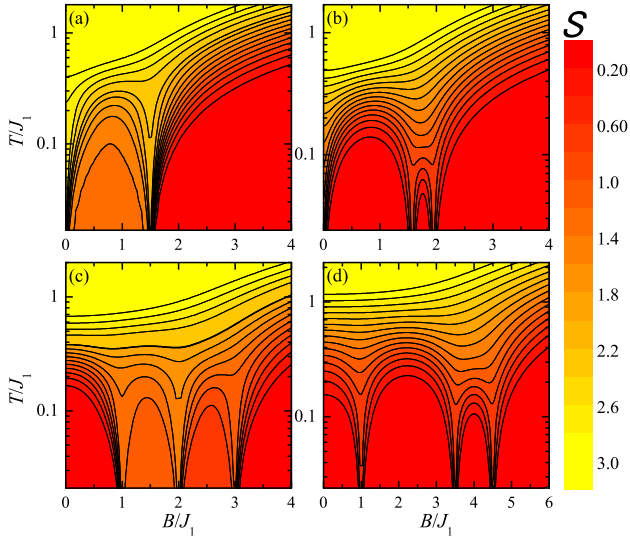}
\caption{\label{fig:8} (color online) Density plot of the entropy, in a color gradient scale
 shown to the right, as a function the magnetic field $B/J_1$ and the temperature $T/J_1$ (in
 logarithmic scale). Some lines of constant entropy are also plotted. In (a) to (d) the exchange
 interactions are the same as in Fig. \ref{fig:5}.}
\end{figure}

Another way to analyze the global entropic behavior of this system is to look at the entropy
as a function of the external magnetic field and the temperature, as is illustrated in a
gradient color scale in Fig. \ref{fig:8} for various values of the exchange interactions.
To better visualize the topology of the surface, some lines of constant entropy have been
added to the figure. These lines illustrate, basically, the isentropic variation of temperature
with the external magnetic field. The spikes of the constant entropy lines at low temperatures are in
fact related to the transitions at $T=0$ and are located at the corresponding critical
external magnetic fields. It can also be clearly seen that as the intertriangle interaction
increases the degeneracy is systematically broken, specially at zero magnetic field. The
magnetocaloric effect, enhanced during the adiabatic (de) magnetization, can be found at
zero magnetic field in Figs. \ref{fig:8}(a) and (b). This effect will be discussed in more
details in the next subsection.

\subsection{Magnetocaloric effect}
\label{mce}

Relevant to the study of the magnetocaloric effect, which is in some way implicit in
the results shown in Figs.
\ref{fig:7} and \ref{fig:8}, is the isothermal change of the entropy with the variation of
the external magnetic field. Such a quantity, basically related to the numerator of Eq.
\eqref{mceq}, can be given by
\begin{eqnarray}
  \Delta \mathcal{S}(B,T)=\mathcal{S}(B_f,T)-\mathcal{S}(B_i,T),
\end{eqnarray}
where $B_i$ and $B_f$ are, respectively, the initial and final magnetic field.
\begin{figure}[h]
\includegraphics[scale=0.7]{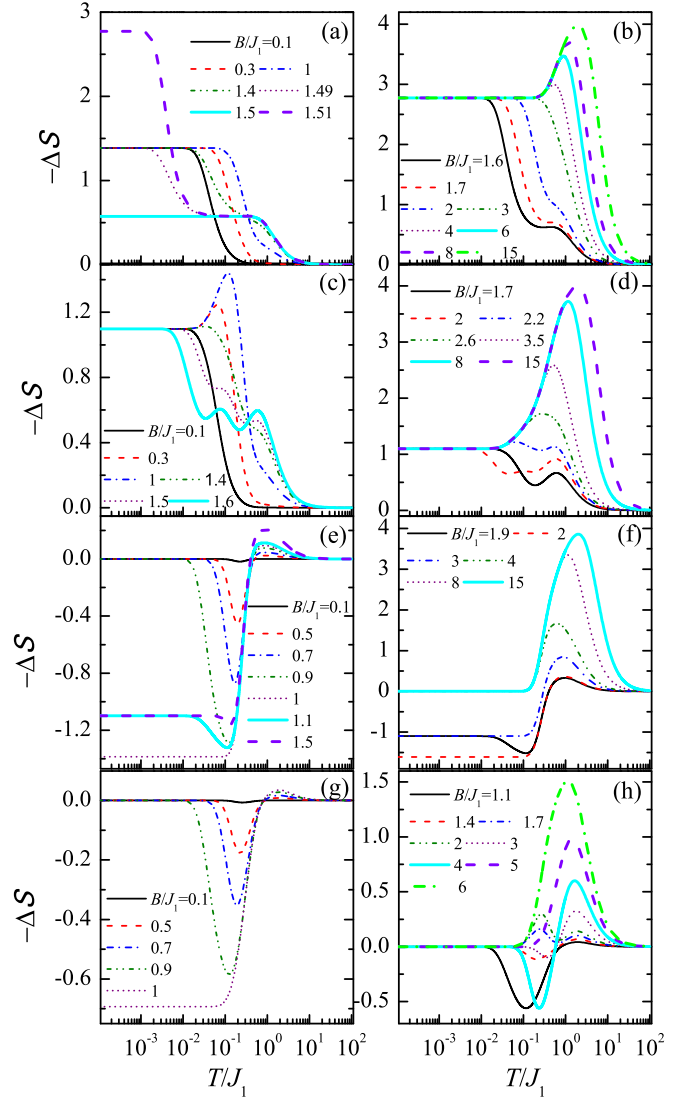} \caption{\label{fig:9} (color online) Isothermal change
 of the entropy $-\Delta \mathcal{S}$, as a function of the temperature $T/J_1$ (in logarithmic
 scale), for several values of the final magnetic field $B/J_1$.
 (a) and (b) for $J_2/J_1=1$ and $J_3/J_1=0$; (c) and (d) for $J_2/J_1=0.5$ and $J_3/J_1=0.3$;
 (e) and (f) for $J_2/J_1=1$ and $J_3/J_1=1$; (g) and (h) for $J_2/J_1=0.5$ and $J_3/J_1=2$.}
\end{figure}
Although it is not necessarily the initial reference field being zero, we will take
$B_i=0$ in all the following results. Fig. \ref{fig:9} shows $-\Delta \mathcal{S}$ as a
function of temperature $T/J_1$ for various final magnetic field values. We are plotting
the negative of the isothermal entropy change because most experimental data have been
reported in this way. For all values of the exchange parameters,
$\Delta \mathcal{S}=0$ at high temperatures for any magnetic field, since in this case all
entropies have the same value, as can be seen in Fig. \ref{fig:7}. In Figs. \ref{fig:9}(a)
and (b) we have the results for uncoupled triangles ($J_3=0$). We can see that in this
case  $-\Delta \mathcal{S}$ is always positive, reflecting the fact that at $B=0$ the
entropy has the greatest value, as can be clearly seen in Fig. \ref{fig:7}(a).
For values of the final magnetic field close to the critical field the curves show two
plateaus with increasing the temperature.
Above the critical field, the plateaux start developing a shoulder, that eventually transform
in peaks that become higher and broader with the increase of the magnetic field.

For a small coupling between the triangles ($J_3/J_1=0.3$), as depicted in Figs. \ref{fig:9}(c)
and (d), the two plateaux are suppressed. As the final magnetic field increases from zero,
a peak initially grows up. However, after a certain value of the magnetic field, the peak starts
to decrease and becomes a double shoulder. For final fields close to the
critical fields, and in the region between them, a double peak structure forms, and
by still increasing the final fields one ends up with just one peak again, which becomes
higher and broader. This double peak structure, for some values of the final magnetic
field, has already been observed in the theoretical treatment of ref.  \cite{kowa} which
took into account only the estimated experimental coupling constants of V$_6$ magnetic molecules.

It is still more interesting the behavior of the magnetocaloric effect for higher values
of the intertriangle interaction, which are shown in Figs. \ref{fig:9} (e)-(f) for
$J_3=1$ and in Figs. \ref{fig:9} (g)-(f) for $J_3=2$. In these cases negative values for
entropy change is obtained, including negative valleys and positive peaks.

Generally speaking, the peaks in Fig. \ref{fig:9} correspond to the called direct magnetocaloric
effect, while the valleys correspond to the inverse magnetocaloric effect (see ref. \cite{franco}
and references therein).

We have also considered negative values of the exchange interactions and the
behavior of the magnetocaloric effect has no qualitative changes compared to the results
presented above. 

\subsection{Specific heat}
\label{sph}

\begin{figure}[h]
\includegraphics[scale=0.45]{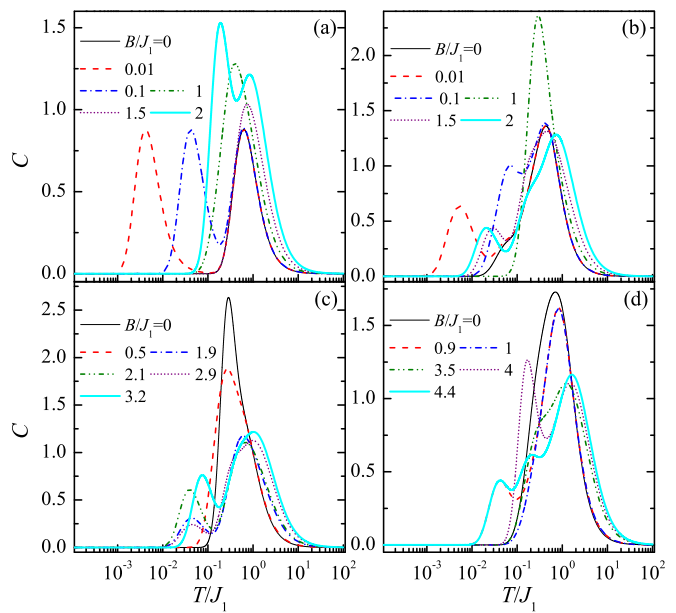} \caption{\label{fig:10} (color online) Specific heat $C$
 as a function of temperature $T/J_1$ (in logarithmic scale) for various values of the magnetic
 field $B/J_1$. The exchange interactions in (a)-(d) are the same as those given in Fig.
 \ref{fig:5}.}
\end{figure}

For completeness, it is also instructive to discuss the specific heat behavior of these
molecular magnets. Since the
energy spectrum of the molecules are finite, the specific heat should present a kind of
Schottky behavior in the sense that it should be zero as the temperature goes either to
zero or infinity. This is indeed the fact, as is depicted in Fig. \ref{fig:10}, where the
specific heat $C$ has been plotted as a function of temperature  $T/J_1$ (in logarithmic
scale) for various values of the magnetic field and exchange interactions.
The usual
Schottky specific heat behavior has only one peak, due to fact that the entropy is a
monotonic increasing function of the temperature. At low temperatures, the entropy is
close to zero (or to its residual value) with zero derivative, while at high temperatures
all states are equally probable and the entropy reaches its temperature saturation value also
with zero derivative.  However, looking at the entropy behavior shown in Fig. \ref{fig:7}
we can see that in some instances $\mathcal{S}$ increases in two steps, with a plateau in
between. Each side of the plateau will then contribute to a peak in the specific heat.

\begin{figure}[h]
\includegraphics[scale=0.4]{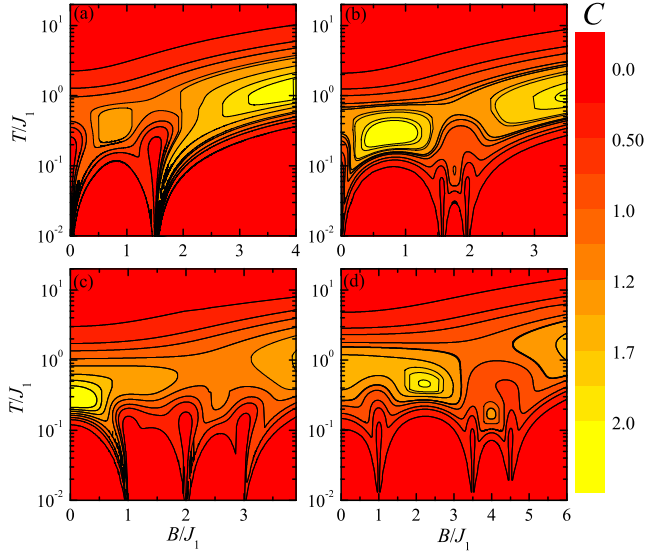}\\
\vspace{0.7cm}
\caption{\label{fig:11} Specific heat $C$, in a color gradient scale shown to the right, as a
 function of temperature T/J1 (in logarithmic scale). Some lines of constant $C$ are also plotted.
 In (a) to (d) the exchange interactions are the same as in Fig. \ref{fig:5}. }
\end{figure}

In Fig.\ref{fig:11} we have the global specific heat $C$, in a color gradient scale, as a
function of temperature $T/J_1$ (in logarithmic scale) and magnetic field. The lines are
constant values of $C$ to better visualize the topology of the function.
Here we can observe that for intermediate temperatures there is a maximum of specific heat
(light color), and we also see that there are peaks of the specific heat near the critical
magnetic field at low temperatures.

\section{Concluding remarks}
\label{conc}

A quantum spin-1/2 Heisenberg model, composed of two coupled isosceles
triangles with an external magnetic field applied along the z-axis, has been
theoretically analyzed through the exact diagonalization of the Hamiltonian
matrix. This model is suitable to describe V$_6$ magnetic molecules and has been
previously studied only for small values of some exchange interactions that come
from fits to the experimental realizations of polyoxovanadates compounds. We have
here considered a wider range of the Hamiltonian parameters for different values of
the external magnetic field.

We have studied the behavior of the system at zero temperature through its energy spectrum,
making a detailed analysis of the corresponding phase diagrams for different coupling values.
In addition, we have shown that the system presents not only a rich phase diagram, but also
residual entropies at zero temperature. We have as well made a detailed study of the magnetic
and thermodynamic properties of the system as a function of temperature. Magnetization,
susceptibility, entropy, magnetocaloric effect and specific heat have been computed. In
some limiting cases, the model makes indeed an excellent qualitative description of the V$_6$
magnetic molecules in comparison with the experimental results obtained in references
\cite{luban,haraldsen}.

 It is interesting that the model (\ref{ham}), being a quite simple example
of a zero-dimensional quantum system, exhibits a very rich thermodynamic behavior and can be
applied not only to generalizations of the V$_6$ molecules but also to low dimensional quantum
models such as triangular chains.

\begin{acknowledgments}
The authors would like to thank CNPq (JT 159792/2019-3), Capes and FAPEMIG for financial
support.
\end{acknowledgments}

\appendix*
\section{}

Hamiltonian \eqref{ham} is represented by a $64 \times 64$ matrix and can be exactly
diagonalized using, for example, Maplesoft\texttrademark~ software. The eigenvalues
$\epsilon_i$, with $1\le i \le64$, can be written as
\begingroup
\allowdisplaybreaks
\begin{align*}
  \varepsilon_1=&3B+J_1+\dfrac{1}{2}J_2+\dfrac{3}{2}J_3,\\
  \varepsilon_2=&-3B+J_1+\dfrac{1}{2}J_2+\dfrac{3}{2}J_3,\\
  \varepsilon_3=&2B+\dfrac{1}{2}J_1-\dfrac{1}{2}J_2,\\
  \varepsilon_4=&-2B+\dfrac{1}{2}J_1-\dfrac{1}{2}J_2,\\
  \varepsilon_5=&2B-\dfrac{1}{2}J_1+\dfrac{1}{2}J_2,\\
  \varepsilon_6=&-2B-\dfrac{1}{2}J_1+\dfrac{1}{2}J_2,\\
  \varepsilon_7=&2B+\dfrac{1}{2}J_1-\dfrac{1}{2}J_2+J_3,\\
  \varepsilon_8=&-2B+\dfrac{1}{2}J_1-\dfrac{1}{2}J_2+J_3,\\
  \varepsilon_9=&2B-\dfrac{1}{2}J_1+\dfrac{1}{2}J_2+J_3,\\
  \varepsilon_{10}=&-2B-\dfrac{1}{2}J_1+\dfrac{1}{2}J_2+J_3,\\
  \varepsilon_{11}=&2B+J_1+\dfrac{1}{2}J_2+\dfrac{3}{2}J_3,\\
 \varepsilon_{12}=&-2B+J_1+\dfrac{1}{2}J_2+\dfrac{3}{2}J_3,\\
  \varepsilon_{13}=&2B+J_1+\dfrac{1}{2}J_2-\dfrac{1}{2}J_3,\\
  \varepsilon_{14}=&-2B+J_1+\dfrac{1}{2}J_2-\dfrac{1}{2}J_3,\\
  \varepsilon_{15}=&B+\dfrac{1}{2}J_1-\dfrac{1}{2}J_2,\\
  \varepsilon_{16}=&-B+\dfrac{1}{2}J_1-\dfrac{1}{2}J_2,\\
  \varepsilon_{17}=&B-\dfrac{1}{2}J_1+\dfrac{1}{2}J_2,\\
  \varepsilon_{18}=&-B-\dfrac{1}{2}J_1+\dfrac{1}{2}J_2,\\
  \varepsilon_{19}=&B+\dfrac{1}{2}J_1-\dfrac{1}{2}J_2+J_3,\\
  \varepsilon_{20}=&-B+\dfrac{1}{2}J_1-\dfrac{1}{2}J_2+J_3,\\
  \varepsilon_{21}=&B-\dfrac{1}{2}J_1+\dfrac{1}{2}J_2+J_3,\\
  \varepsilon_{22}=&-B-\dfrac{1}{2}J_1+\dfrac{1}{2}J_2+J_3,\\
  \varepsilon_{23}=&B+\dfrac{1}{2}J_1-\dfrac{1}{2}J_2-J_3,\\
  \varepsilon_{24}=&-B+\dfrac{1}{2}J_1-\dfrac{1}{2}J_2-J_3,\\
  \varepsilon_{25}=&B-\dfrac{1}{2}J_1+\dfrac{1}{2}J_2-J_3,\\
  \varepsilon_{26}=&-B-\dfrac{1}{2}J_1+\dfrac{1}{2}J_2-J_3,\\
  \varepsilon_{27}=&B+J_1+\dfrac{1}{2}J_2+\dfrac{3}{2}J_3,\\
  \varepsilon_{28}=&-B+J_1+\dfrac{1}{2}J_2+\dfrac{3}{2}J_3,\\
  \varepsilon_{29}=&B+J_1+\dfrac{1}{2}J_2-\dfrac{1}{2}J_3,\\
  \varepsilon_{30}=&-B+J_1+\dfrac{1}{2}J_2-\dfrac{1}{2}J_3,\\
  \varepsilon_{31}=&B-J_1-\dfrac{1}{2}J_2+\dfrac{1}{2}J_3,\\
  \varepsilon_{32}=&-B-J_1-\dfrac{1}{2}J_2+\dfrac{1}{2}J_3,\\
  \varepsilon_{33}=&B-\dfrac{1}{4}J_1-\dfrac{1}{2}J_2-\frac{1}{4}J_3
  +  \dfrac{1}{4}p,\\
  \varepsilon_{34}=&-B-\dfrac{1}{4}J_1-\dfrac{1}{2}J_2-\dfrac{1}{4}J_3
  +  \dfrac{1}{4}p,\\
  \varepsilon_{35}=&B-\dfrac{1}{4}J_1-\dfrac{1}{2}J_2-\dfrac{1}{4}J_3
  -  \dfrac{1}{4}p,\\
  \varepsilon_{36}=&-B-\dfrac{1}{4}J_1-\dfrac{1}{2}J_2-\dfrac{1}{4}J_3
  -  \dfrac{1}{4}p,\\
  \varepsilon_{37}=&B-\dfrac{5}{4}J_1+\dfrac{1}{2}J_2-\dfrac{1}{4}J_3+  \dfrac{1}{4}p,\\
  \varepsilon_{38}=&-B-\dfrac{5}{4}J_1+\dfrac{1}{2}J_2-\dfrac{1}{4}J_3
  +   \dfrac{1}{4}p,\\
  \varepsilon_{39}=&B+\dfrac{3}{4}J_1-\dfrac{3}{2}J_2-\dfrac{1}{4}J_3
  -  \dfrac{1}{4}p,\\
  \varepsilon_{40}=&-B+\dfrac{3}{4}J_1-\dfrac{3}{2}J_2-\dfrac{1}{4}J_3
  -  \dfrac{1}{4}p,\\
  \varepsilon_{41}=&B-\dfrac{1}{2}J_1+\dfrac{1}{2}J_2-J_3
  +  \dfrac{1}{2}q,\\
  \varepsilon_{42}=&-B-\dfrac{1}{2}J_1+\dfrac{1}{2}J_2-J_3
  +  \dfrac{1}{2}q,\\
  \varepsilon_{43}=&B-\dfrac{1}{2}J_1+\dfrac{1}{2}J_2-J_3
  -  \dfrac{1}{2}q,\\
  \varepsilon_{44}=&-B-\dfrac{1}{2}J_1+\dfrac{1}{2}J_2-J_3
  -  \dfrac{1}{2}q,\\
  \varepsilon_{45}=&\dfrac{1}{2}J_1-\dfrac{1}{2}J_2,\\
  \varepsilon_{46}=&-\dfrac{1}{2}J_1+\dfrac{1}{2}J_2,\\
  \varepsilon_{47}=&\dfrac{1}{2}J_1-\dfrac{1}{2}J_2+J_3,\\
  \varepsilon_{48}=&-\dfrac{1}{2}J_1+\dfrac{1}{2}J_2+J_3,\\
  \varepsilon_{49}=&\dfrac{1}{2}J_1-\dfrac{1}{2}J_2-J_3,\\
  \varepsilon_{50}=&-\dfrac{1}{2}J_1+\dfrac{1}{2}J_2-J_3,\\
  \varepsilon_{51}=&J_1+\dfrac{1}{2}J_2-\dfrac{1}{2}J_3,\\
  \varepsilon_{52}=&-J_1-\dfrac{1}{2}J_2-\dfrac{1}{2}J_3,\\
  \varepsilon_{53}=&-J_1-\dfrac{1}{2}J_2+\dfrac{1}{2}J_3,\\
  \varepsilon_{54}=&J_1+\dfrac{1}{2}J_2+\dfrac{3}{2}J_3,\\
  \varepsilon_{55}=&-J_1-\dfrac{1}{2}J_2-\dfrac{3}{2}J_3,\\
  \varepsilon_{56}=&-\dfrac{3}{2}J_2-\dfrac{1}{2}J_3,\\
  \varepsilon_{57}=&-\dfrac{1}{4}J_1-\dfrac{1}{2}J_2-\dfrac{1}{4}J_3
  +\dfrac{1}{4}p,\\
  \varepsilon_{58}=&-\dfrac{1}{4}J_1-\dfrac{1}{2}J_2-\dfrac{1}{4}J_3
  -  \dfrac{1}{4}p,\\
  \varepsilon_{59}=&-\dfrac{5}{4}J_1+\dfrac{1}{2}J_2-\dfrac{1}{4}J_3
  +  \dfrac{1}{4}p,\\
  \varepsilon_{60}=&\dfrac{3}{4}J_1-\dfrac{3}{2}J_2-\dfrac{1}{4}J_3
  -  \dfrac{1}{4}p,\\
  \varepsilon_{61}=&-\dfrac{1}{2}J_1+\dfrac{1}{2}J_2+J_3
  +  \dfrac{1}{2}q,\\
  \varepsilon_{62}=&-\dfrac{1}{2}J_1+\dfrac{1}{2}J_2-J_3
  -  \dfrac{1}{2}q,\\
  \varepsilon_{63}=&-\dfrac{1}{2}J_1+\dfrac{1}{2}J_2-J_3
  +  \dfrac{1}{2}r,\\
  \varepsilon_{64}=&-\dfrac{1}{2}J_1+\dfrac{1}{2}J_2-J_3
  -  \dfrac{1}{2}r,
\end{align*}
\endgroup
where
\begin{align*}
  p=&\sqrt{9J_1^2-10J_1J_3+17J_3^2},\\
  q=&\sqrt{9J_1^2-10J_1J_3+5J_3^2},\\
  r=&\sqrt{9J_1^2-18J_1J_3+13J_3^2}.
\end{align*}

The corresponding eigenstates can also be obtained, but as they are rather lengthy they
are not reproduced here. Only the most relevant ones for the quantum phase diagrams are
given in the text.

\end{document}